\documentclass[conference]{IEEEtran}

\IEEEoverridecommandlockouts                              

\usepackage{cite}
\usepackage{amsmath,amssymb,amsfonts}
\usepackage[dvipsnames]{xcolor}
\usepackage{acro}
\usepackage{subcaption}
\usepackage{pgf}
\usepackage{makecell}

\usepackage{amsthm}
\usepackage[binary-units]{siunitx}[=v2]
\usepackage{mathtools}
\usepackage{pgfplots}
\pgfplotsset{compat=1.8}
\usepgfplotslibrary{statistics}
\usepackage{tikz}
\usepackage{pdfpages}
\usepackage[T1]{fontenc}
\usepackage{bm}
\usepackage{forest}
\usepackage{floatrow}
\usepackage{comment}
\usepackage{bm}
\usepackage{xfrac}
\usepackage{tikz}
\usepackage{colortbl}
\usepackage{xcolor}

\PassOptionsToPackage{svgnames}{xcolor}
\usepackage{tcolorbox}
\tcbuselibrary{skins,breakable}
\usetikzlibrary{shadings,shadows}

    {\endtcolorbox}

\title{Generative AI in Systems Engineering: A Framework for Risk Assessment of Large Language Models}

\author{\IEEEauthorblockN{Stefan Otten\IEEEauthorrefmark{1}, Philipp Reis\IEEEauthorrefmark{1}, Philipp Rigoll\IEEEauthorrefmark{1}, Joshua Ransiek\IEEEauthorrefmark{1},
Tobias Schürmann\IEEEauthorrefmark{1},
Jacob Langner\IEEEauthorrefmark{1}, and Eric Sax\IEEEauthorrefmark{1}}
\IEEEauthorblockA{FZI Research Center for Information Technology,
Karlsruhe, Germany\\
Email: \IEEEauthorrefmark{1}otten,reis,rigoll,ransiek,langner,schuermann,sax@fzi.de}
}

\begin{document}
\theoremstyle{definition}
\newtheorem{definition}{Definition}
\renewcommand{\theadalign}{vh}
\newcommand{\probP}{\text{I\kern-0.15em P}}

\newcommand{\mathdefault}[1][]{}

\maketitle
\pagestyle{empty}


\begin{abstract}

The increasing use of Large Language Models (LLMs) offers significant opportunities across the engineering lifecycle, including requirements engineering, software development, process optimization, and decision support. Despite this potential, organizations face substantial challenges in assessing the risks associated with LLM use, resulting in inconsistent integration, unknown failure modes, and limited scalability.

This paper introduces the LLM Risk Assessment Framework (LRF), a structured approach for evaluating the application of LLMs within Systems Engineering (SE) environments. The framework classifies LLM-based applications along two fundamental dimensions: autonomy, ranging from supportive assistance to fully automated decision making, and impact, reflecting the potential severity of incorrect or misleading model outputs on engineering processes and system elements. By combining these dimensions, the LRF enables consistent determination of corresponding risk levels across the development lifecycle.

The resulting classification supports organizations in identifying appropriate validation strategies, levels of human oversight, and required countermeasures to ensure safe and transparent deployment. The framework thereby helps align the rapid evolution of AI technologies with established engineering principles of reliability, traceability, and controlled process integration. Overall, the LRF provides a basis for risk-aware adoption of LLMs in complex engineering environments and represents a first step toward standardized AI assurance practices in systems engineering.

\end{abstract}

\begin{IEEEkeywords}
Large Language Models, Systems Engineering, Risk Framework, Generative AI
\end{IEEEkeywords}

\section{Introduction}
\label{sec:introduction}

\begin{figure}
\centering
\includegraphics[width=\textwidth]{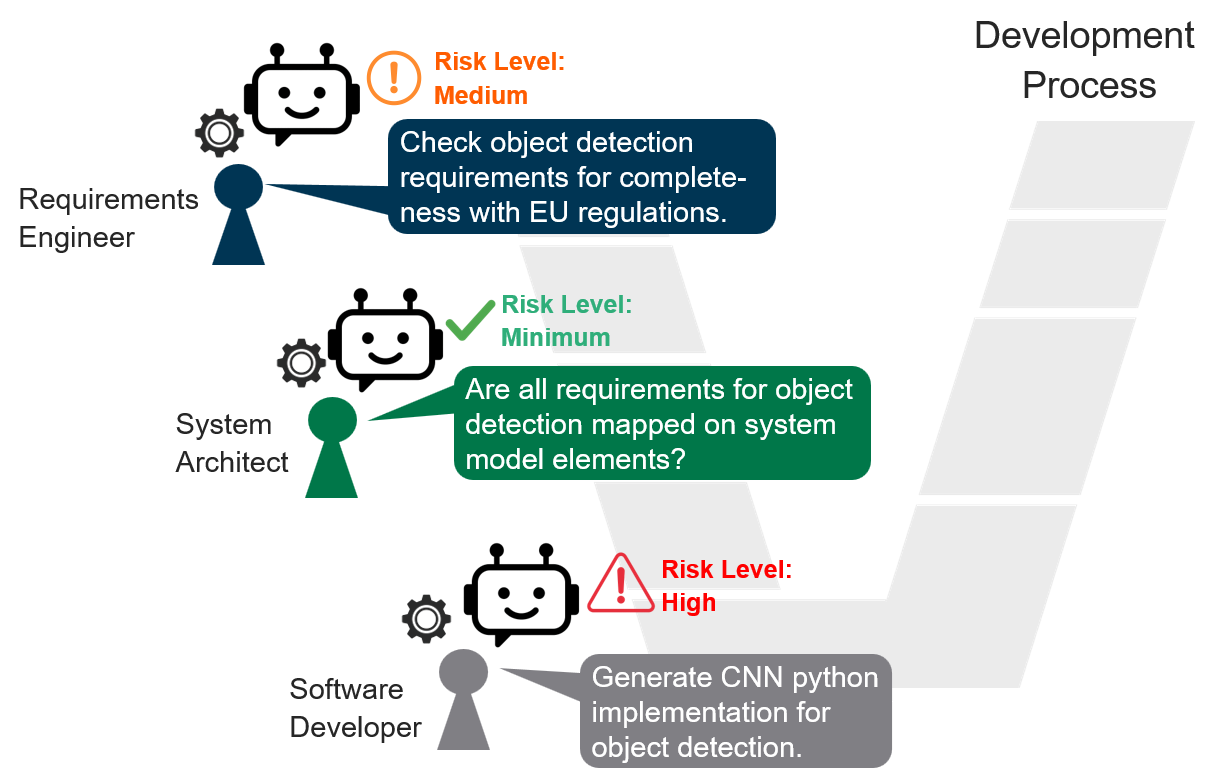}%
\caption{Potential use and arising risks of LLMs in systems engineering}
\label{fig:llmSE}
\end{figure}

Generative artificial intelligence solutions, driven particularly by LLMs, are becoming increasingly present in professional environments~\cite{Feuerriegel2024}. Their widespread availability and intuitive use have made them an integral part of daily work and communication. In industrial and engineering contexts, LLMs are used in a variety of applications~\cite{Hovemann2025}, although they are often isolated and separated in established business processes or systems engineering tasks~(see Fig. \ref{fig:llmSE}). While these implementations demonstrate the disruptive potential of LLMs, they frequently lack systematic integration into the structured lifecycle of cyber-physical systems~\cite{Hohl2025} and organisational structure.

The pace of technological progress in this field contrasts with the stability, quality orientation and risk awareness typically required in the development of these systems. Engineering processes such as the V-Model~\cite{droschel_v-modell_2000} are established for structured and systematic development, whereas LLMs evolve rapidly and in less predictable ways. As a result, organizations face difficulties in aligning the agility of AI technologies with the rigor of established engineering lifecycles. This creates uncertainty about reliability and potential risks of using LLM based solutions in engineering environments and requires systematic and methodological approaches~\cite{Petersen2022}.

To address these challenges, organizations require a structured framework that enables a quick but systematic assessment of risks arising from LLM adoption in engineering. Such an approach supports flexible and safe integration of LLM applications into existing processes while maintaining the reliability and transparency expected in engineering disciplines. Moreover, the implementation of a comprehensive framework promotes consistent evaluation and standardization across the development process and projects, ensuring a unified approach to LLM deployment throughout organizations. The LLM Risk Assessment Framework~(LRF) introduced in this paper aims to provide this foundation by classifying LLM use cases according to their autonomy and impact, thereby enabling consistent risk evaluation and controlled deployment.

\section{Related Work}\label{sec:related_work}

Research on LLMs in engineering consistently highlights reliability, alignment, and lifecycle risks. Hallucinations, probabilistic behavior, bias, and limited interpretability complicate verification, requirements satisfaction, and safe integration into socio-technical systems, underscoring the need for established systems engineering practices to manage maintainability, security, and privacy \cite{cabrera_systems_2024,chen_systems_2024}.

Beyond these risk-oriented perspectives, recent work emphasizes the need to embed LLMs more deliberately within engineering methodologies. LLMs have been interpreted as extensions of established optimization and modeling practices, clarifying their underlying assumptions and operational characteristics while highlighting their potential to support broader engineering workflows \cite{patterson_large_nodate}.

\begin{figure*}
\centering
\includegraphics[width=\textwidth]{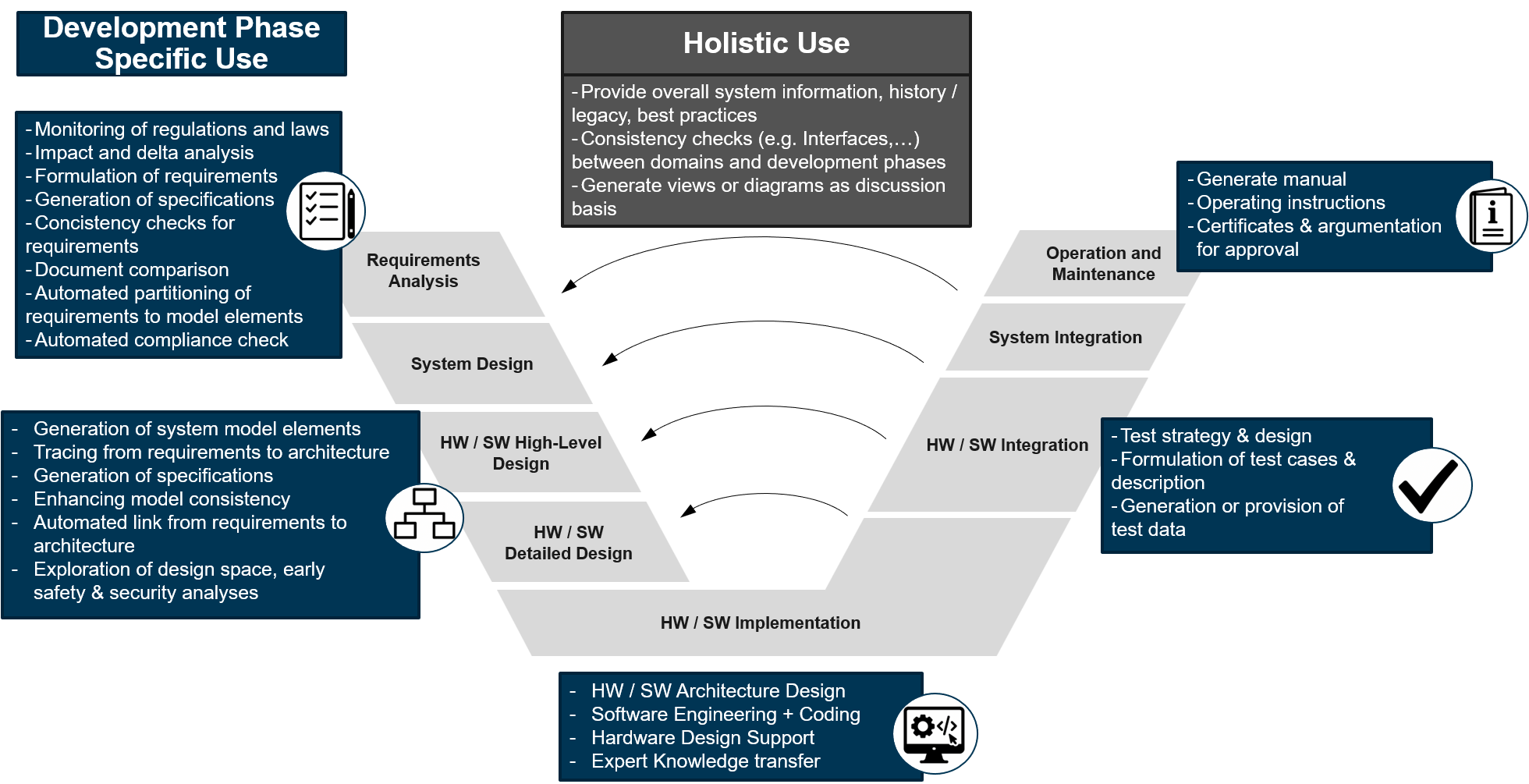}%
\caption{Areas of application of LLMs in the development process}
\label{fig:vmodell}
\end{figure*}

Such capabilities increasingly span coordination, reporting, compliance-related activities, and decision-support functions within design processes. Complementary research emphasizes that such expanded roles require clearer specification practices, structured outputs, and process supervision to mitigate the ambiguity inherent in natural-language interfaces and to enable more reliable LLM-based components \cite{stoica_specifications_2024}.

Together, these contributions highlight the growing expectation that LLMs will participate in higher-level engineering tasks, while leaving open the question of how to assess the risk of such involvement.

Applied studies demonstrate growing use of LLMs across engineering workflows, including knowledge access, collaboration, code analysis, documentation processing, and virtual assistance. Adoption guidance highlights the importance of integration, data governance, customization, and training \cite{Fallmann_how_24}. Domain-specific work in model-based systems engineering shows that LLMs can generate system architectures and components with reasonable alignment to requirements, though human oversight remains essential \cite{timperley_assessment_2025}. Enterprise analyses further explore lifecycle models, customization techniques, infrastructure demands, and governance considerations for large-scale deployment \cite{singh_systems_2025}. Despite the breadth of these applications, they are rarely classified in terms of autonomy or system impact.

Overall, existing literature offers valuable insights into challenges, engineering requirements, and emerging applications, but it does not provide a unified means to assess the risk of LLM use across domains. The LLM Risk Assessment Framework introduced in this paper addresses this gap by structuring applications along autonomy and system impact, enabling consistent and transparent risk assessment for LLM integration in systems engineering.

This paper makes the following three contributions

\begin{itemize}
    \item Propose the \textbf{LLM Risk Assessment Framework }(LRF): A structured, domain-agnostic model that classifies LLM applications in systems engineering along autonomy level and system impact.

    \item Enable \textbf{systematic risk assessment}: Link autonomy levels and impact of the LLM to corresponding risk levels, providing a consistent method to evaluate and compare heterogeneous LLM use cases.
    
    \item Support \textbf{safe and scalable integration}: Derive suitable mitigation measures for each development, guiding organizations in reliable and transparent LLM adoption.
\end{itemize}
\section{Application of LLMs in the Systems Engineering Processes}

LLMs can potentially support, improve, and accelerate the engineering lifecycle in a wide range of activities (see Fig.~\ref{fig:vmodell}). Their ability to process natural language, reason across heterogeneous information, and generate coherent technical content allows them to act as intelligent assistants across different engineering roles and process phases. From early concept definition to system verification and release, LLMs can provide both broad and specific contributions that enhance collaboration, consistency, and efficiency within systems engineering processes.

\subsection{Holistic Engineering Use}

From a cross-disciplinary perspective, LLMs offer comprehensive assistance that spans across engineering domains, project roles, and viewpoints. They can act as interactive system companions that provide overviews of complex architectures, dependencies, and design rationales. Through integration with engineering repositories, LLMs can establish a knowledge base that supports knowledge transfer across teams and project stages. Moreover, their analytical capabilities enable engineers to identify constraints and inconsistencies, such as missing links between requirements, conflicting design assumptions, or unconsidered regulatory obligations. LLMs can assist in defining the operational design domain of systems and identifying relevant standards, laws, and regulations to ensure compliance early in the development process.

\subsection{Development Phase-Specific Use}

Within system development, LLMs can be applied for specific tasks tailored to different individual development phases.

\textit{Requirements Engineering:}
LLMs can support the derivation and generation of requirements~\cite{Bazzal2025} from system descriptions, use cases, or regulatory texts. They can verify completeness and consistency, detect contradictions, and trace the relationship between requirements and applicable standards. Furthermore, they can generate summaries and overviews that help maintain transparency across stakeholder perspectives.

\textit{System and Architecture Design:}
LLMs can assist designing and engineering the system~\cite{Hovemann2025,Crabb2024} and maintaining consistency of the development process by linking artifacts and providing natural language access to system models, design documents, and requirement sets. Their ability to summarize, query, and reason across data enables a continuous documentation process and facilitates design reviews.

\textit{Verification and Validation:}
During testing and integration, LLMs can automate parts of test design and execution by generating test cases, scripts, and data, and by evaluating test completeness and coverage. They can link requirements with corresponding test cases, generate documentation, and assist in the interpretation of test results to identify potential weaknesses or inconsistencies. Automated simulation models for virtual evaluation of system properties are also investigated~\cite{Zhang2025}.

\textit{Release and Operation:}
In the final development phases, LLMs can support the release process by ensuring the completeness and consistency of documentation and user manuals. They can automatically generate or update manuals, check alignment with requirements and regulatory needs, and aggregate evidence for compliance and release readiness.

The broad range of potential applications demonstrates the transformative and disruptive potential of LLMs in engineering.

\section{Challenges using LLMs in Systems Engineering}

The increasing integration of LLMs into systems engineering processes is accompanied by a variety of challenges that must be considered to ensure their reliable and responsible use. These challenges can be grouped into two main categories: technical and model-related limitations and human and organizational factors.

\subsection{Technical and Model-Related Challenges}

From a technical perspective, LLMs exhibit several inherent limitations that directly affect their applicability in engineering contexts~\cite{kaddour2023challenges}. One major constraint involves implementation and inference costs, as large models require substantial computational resources and memory capacity. 

A further restriction arises from the limited context length, which hinders the model’s ability to handle extended input sequences such as detailed specifications, documentation, or system descriptions. This limitation becomes a barrier for tasks that require reasoning across excessive or interdependent documents or models. Techniques such as Retrieval Augmented Generation (RAG) adress these restrictions~\cite{graphrag2025,Wang2024}.

Another key challenge is prompt brittleness, meaning that small syntactic or semantic variations in prompts can cause substantial changes in the model output, often occuring unintuitive to humans. This behavior complicates the development of stable and repeatable LLM based workflows in engineering. Similarly, hallucination remains a fundamental risk: LLMs can generate fluent but factually incorrect or unfounded statements, potentially compromising design integrity, documentation, or compliance information if undetected.

The need for systematic validation and evaluation further complicates the deployment of LLMs. Current benchmarks are often brittle and static, with results that can vary widely depending on the prompt formulation or evaluation protocol. Moreover, most benchmarks are manually curated and not aligned with specific engineering tasks. This lack of controlled experimentation and standardized validation makes it difficult to assess model reliability and maturity in practical use cases.

Finally, explainability and traceability represent key concerns. Within systems engineering, every design decision must be traceable and justified. However, LLMs provide limited transparency regarding their internal reasoning, reference sources, or decision paths.

\begin{figure*}
\centering
\includegraphics[width=.9\textwidth]{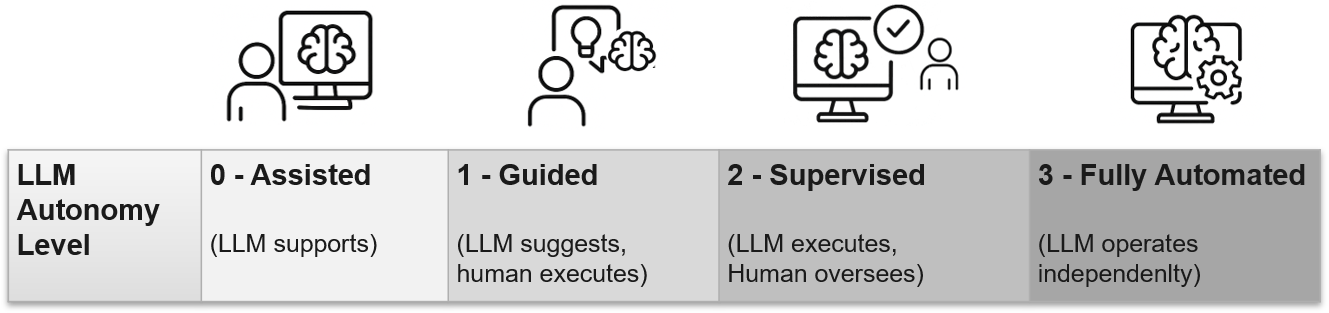}%
\caption{Proposed Autonomy Levels for usage of LLMs}
\label{fig:autonomy_levels}
\end{figure*}

\subsection{Human and Organizational Challenges}

Beyond technical aspects, the human and organizational dimension plays a crucial role in the responsible use of LLMs. One emerging issue is overreliance on LLM generated content~\cite{bucinca_trust_2021}. The continuous use of automated generation tools can lead to a loss of domain knowledge and critical thinking among engineers. Users may adopt model outputs without sufficient verification or understanding, increasing the risk of propagating errors or biases.

Moreover, the difficulty of identifying LLM generated outputs introduces uncertainty into collaborative environments. When human and machine generated content become indistinguishable, version tracking, authorship, and accountability are increasingly difficult to maintain and provides potential for unintentional revealing of confidential data, also named disclosure.

Organizationally, the integration of LLMs requires new governance structures, clear policies on model usage, and well defined responsibilities across teams. The lack of company wide standards for prompt design, model validation, or data management can result in inconsistent practices and fragmented adoption across departments. In addition, many organizations lack data-driven development~\cite{Bach2017} practices required for effective AI integration, leading to inconsistent data availability, quality, and governance across teams.

The challenges outlined above highlight the need for a structured approach to evaluate and manage the risks associated with the use of LLMs in Systems Engineering.

\section{LLM Risk Assessment Framework}\label{sec:Framework}

To enable a structured evaluation of the potential risks associated with the use of LLMs in Systems Engineering, this paper introduces the LLM Risk Assessment Framework~(LRF). The framework aims to support organizations in systematically assessing and classifying LLM-based applications by combining two key dimensions: the degree of autonomy and the associated impact on engineering outcomes.

The degree of autonomy reflects how independently the LLM performs a given task, ranging from supportive assistance to full automation. The impact dimension indicates the severity of potential consequences if an LLM output is incorrect, misleading, or misaligned with design intent. By mapping specific use cases within this two-dimensional space, the framework provides a structured foundation for determining the necessary level of control, validation, and oversight during deployment.

\subsection{LLM Autonomy Levels}

The framework's autonomy dimension defines four levels representing increasing degrees of independence in LLM operation (see Fig. \ref{fig:autonomy_levels}). Each level corresponds to a distinct mode of interaction between the human user and the model, reflecting how decision authority is distributed. This concept is inspired by the SAE levels~\cite{SAE3016} of driving automation, where responsibility and control gradually shift from the human to the system.

Analogously, in the lower levels of the LRF (0 - Assisted and 1 - Guided), the human remains in charge, whereas in the higher levels (2 - Supervised and 3 - Fully Automated), the AI system takes the lead, with humans providing oversight or intervention.

\begin{figure*}
\centering
\includegraphics[width=\textwidth]{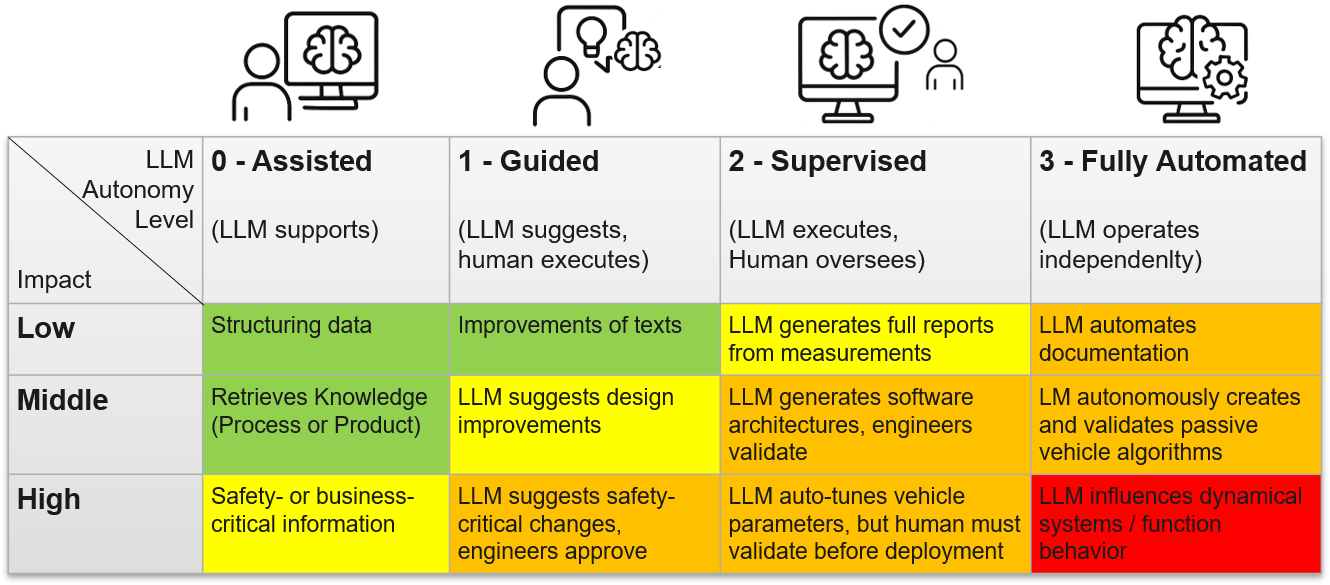}%
\caption{Overview of proposed LLM Risk Assessment Framework (LRF)}
\label{fig:risiko}
\end{figure*}

\textbf{0 - Assisted:}
The LLM provides support to the user without making decisions or initiating actions autonomously. The LLM stays within the context given by the user. The human user retains full control over all outcomes, and the LLM acts purely as a supportive tool.

\textrightarrow{}~Human in charge: the model improves human input.

\textbf{1 - Guided:}
The LLM suggests actions or recommendations that require human confirmation and review before execution. This mode is common for copilots or chat assistants that propose design alternatives, parameter settings, or analytical insights, but where the engineer must explicitly approve each action. Therefore, in comparison to level 0, the LLM explores and generates alternative ideas on its own, out of the given context or input by the user.

\textrightarrow{}~Human in charge: the model guides and recommends, but execution remains under human authority.

\textbf{2 - Supervised:}
The LLM performs defined tasks independently, such as generating requirement sets, producing design documentation, or preparing test cases. Human users monitor the process and intervene when necessary to ensure that outputs remain accurate and contextually correct.

\textrightarrow{}~AI in charge under human supervision: the model executes autonomously, but human oversight is maintained.

\textbf{3 - Fully Automated:}
The LLM operates automated without human oversight, directly performing actions and taking decisions. The output of the LLM is directly used without further supervision by the user, significantly increasing potential risks.

\textrightarrow{}~AI in charge: the model acts independently, executing decisions or generating artifacts without human confirmation.

As with SAE classification levels for automated driving, higher autonomy levels correlate with greater reliance on system reliability, transparency, and validation. The shift from human control to AI control requires increasingly rigorous safety and assurance measures to ensure responsible operation.

\subsection{Impact Level}

The impact dimension of the framework describes the severity of the consequences resulting from incorrect or misleading behavior by the LLM. The classification is derived from the potential influence of LLM output on engineering activities and their downstream effects on product quality, safety, and compliance. The impact level therefore reflects how critical the LLM’s contribution is within the overall systems engineering process.

\textbf{Low Impact:}
LLM outputs have minimal or reversible consequences for system performance, safety, or compliance.

\textbf{Medium Impact:}
The LLM output has indirect impact on the system or engineering tasks. Mistakes at this level can propagate into subsequent steps, affecting cost, performance, safety or quality.

\textbf{High Impact:}
The LLM directly influences system design, behavior or use of sensitive or compliant-relevant data. Failures or inaccuracies at this level can have severe consequences for safety, reliability, or compliance of the product or company. 


Although this classification is tailored to LLM-based applications, the underlying principle aligns closely with established risk assessment methodologies in technical safety standards, such as IEC 61508~\cite{EN61508} for electric / electronic systems or ISO 26262~\cite{ISO26262} as an automotive-specific derivation. In these frameworks, risks are commonly evaluated by the parameters severity, controllability and exposure. Within the LRF, the impact level primarily corresponds to the severity dimension, describing the potential harm caused by an incorrect LLM output. The autonomy level of the framework reflects and refers to controllability, indicating to what extent a human can intervene or correct the model’s actions. Exposure can be interpreted as the frequency or scale of LLM usage across users and processes rather than a property of the technical system itself.

By establishing this conceptual link, the LRF translates proven principles from traditional safety engineering to the evaluation of cognitive systems. It enables organizations to reason about AI-driven risks using familiar constructs while extending the focus from physical system safety to informational reliability and decision integrity.

\subsection{Risk Classification}

Combining the dimensions of autonomy and impact enables a structured risk classification of LLM applications in Systems Engineering. The resulting matrix represents how increasing model autonomy and growing impact severity jointly determine the overall level of operational risk. Each field in this matrix corresponds to a specific combination of autonomy and impact, which can be associated with distinct requirements for validation, explainability, and oversight.

This classification does not aim to restrict innovation, but to guide responsible adoption. It allows organizations to identify the level of quality assurance and governance necessary for a given LLM use case, supporting both safe experimentation and scalable deployment within established engineering environments. The four resulting risk levels can be interpreted as an operational continuum, ranging from unrestricted usage under quality management supervision to high risk applications that demand extensive validation and fallback mechanisms.

\textbf{Minimal Risk: (highlighted in green)}
Applications in this area combine low autonomy with low impact. They typically involve documentation support, or retrieval tasks where errors do not have any impact on the system. These applications can be used out of the box within standard quality management procedures, provided a human remains in the loop for plausibility checks. No additional regulatory or technical restrictions are required.

\textbf{Low Risk: (highlighted in yellow)}
This level applies to applications with moderate autonomy or limited impact. LLMs may operate semi-independently but do not directly influence critical aspects. Continuous monitoring and explainable AI mechanisms are recommended to ensure output transparency and detect potential deviations early. Examples include summarization tools, requirement consistency checks, or design review assistants.

\textbf{Medium Risk: (highlighted in orange)}
Applications in this range typically exhibit medium to high autonomy and moderate to high impact. The model’s outputs affect design or decision processes, but human validation remains possible. Such use cases require explicit validation steps before deployment and should integrate explainable AI features to justify reasoning paths. A human in the loop approach is crucial to confirm correctness and traceability of generated artifacts. Example scenarios include automated requirement generation, design proposal creation, or test case derivation.

\textbf{High Risk: (highlighted in red)}
At this level, LLMs operate with high autonomy and influence critical or safety relevant system aspects. Their outputs directly affect configurations, control logic, or verification results. These applications demand comprehensive validation, fallback mechanisms, and robust explainability to ensure predictable behavior. Human oversight must remain possible at all times, supported by technical safeguards that prevent uncontrolled model actions. Deployment in this category should follow structured assurance processes similar to safety case development in traditional engineering standards.

The classification framework is visualized as a matrix (see Fig. \ref{fig:risiko}), where the horizontal axis represents the LLM autonomy level (Assisted to Fully Automated) and the vertical axis indicates the potential impact (Low to High). The risk level increases diagonally across the matrix as both autonomy and impact rise, emphasizing that risk is not a fixed property of the model itself but emerges from its use context and integration depth.

By classifying each LLM use case within this structured matrix, organizations can achieve a balanced approach that aligns the flexibility of AI with the discipline of Systems Engineering. The framework thus supports both innovation and accountability by linking technological autonomy with process maturity and quality assurance.

\section{Example Applications}

To illustrate the applicability of the proposed LRF, two representative examples are presented. They demonstrate how different levels of autonomy and impact translate into distinct risk classifications and corresponding control measures. These examples highlight how the framework can guide organizations in determining the appropriate level of human oversight, validation, and explainability for each LLM-based use case.

\subsection{Requirements Checker}

The first example is an LLM-based Requirements Checker that assists engineers in evaluating the quality of requirements. The model analyzes whether individual requirements comply with common standards of requirements such as completeness, consistency, and clarity. In addition, the tool provids suggestions for improvement or reformulated alternatives if a requirement does not compoly to the standards.

This application operates on Autonomy Level 1 (Guided).
The LLM performs analytical checks and flags potential issues.
Based on the identified issues, the LLM provides proposals for reformulations.
In order to implement these proposals, the user must approve every change.
Proposed changes to requirements can affect downstream decisions in the development process.
Therefore, the output of the Requirements Checker has a medium impact.

The risk analysis with the LRF therefore leads to the classification: low risk.
The fact that the user is still in charge and the LLM only prepares possible actions ensures continuous monitoring, and potential deviations are detected immediately.

\subsection{Legal Case Assessment}

Another example is an LLM-based system for legal case assessment.
Users with limited legal expertise describe a situation, and the system decides whether a specific law is applicable in this situation.
Due to their limited legal expertise, users cannot verify for themselves whether the LLM's conclusion is correct.

This setup corresponds to Autonomy Level 3 (Fully Automated).
The legal decision of the system is made without human validation or oversight.
In the event that the system fails to recognize that a law applies, this can lead to compliance violations and, as a result, to direct financial loss or reputational damage.
Therefore, the impact level is high.

Accordingly, such a system falls within the high-risk area of the LRF.
Therefore, to mitigate potential harm, such a system must undergo comprehensive validation, and fallback mechanisms and robust explainability must be implemented.
A structured assurance process must be followed during the development of such a system. 
If used operationally, it may be necessary to reintroduce human oversight, for instance through legal expert verification before applying model outputs in real decisions.

\section{Conclusion \& Outlook}
\label{sec:conclusion_and_outlook}

The presented work addressed the growing need for structured methods to evaluate the adoption and risk of LLMs within systems engineering environments. While the rapid evolution of LLMs introduces remarkable opportunities, it also challenges traditional engineering principles of validation, traceability, and process control.

The proposed LLM Risk Assessment Framework (LRF) provides a foundation for transparent and comparable evaluation of LLM-based applications. Rather than prescribing specific approaches or use cases, it establishes a scalable logic that links model autonomy and impact to corresponding levels of risk for all applications of LLMs in engineering. Beyond the assessment of individual use cases, the framework promotes a unified perspective on LLM deployment across organizations. It supports consistent governance, a shared understanding of risk, and the gradual integration of AI-driven tools into established Systems Engineering lifecycles.

Further research is required to validate and refine the framework. At its current stage, the LRF defines how to determine and systematically structure the risk of LLM applications but does not yet provide methods to measure or quantify the maturity of a LLM application. Future work should therefore focus on developing suitable metrics and evaluation approaches to assess LLM maturity levels and track their evolution over time.

In the long term, the framework can contribute to the standardization of AI assurance practices and serve as a bridge between emerging AI governance concepts and established engineering standards. By providing a structured foundation for risk-aware integration, the LRF supports the reliable, transparent, and sustainable adoption of LLM technologies in complex engineering environments.

\section*{Acknowledgment}
A Large Language Model (ChatGPT by OpenAI) was used to improve the grammar, style, and readability of the paper. 
ChatGPT by OpenAI was used to generate the icons in the figures.
No generative contribution was made to the scientific content, and all intellectual input remains the responsibility of the human authors.
This work was supported by the German Federal Ministry for Economic Affairs and Energy (BMWE) within the project RepliCar with grant number 19A23002I.



\bibliographystyle{IEEEtran}
\bibliography{root}

\end{document}